\documentclass[preprint,pre,aps,superscriptaddress,a4paper]{revtex4}

\usepackage[utf8x]{inputenc}
\usepackage{graphicx} 
\usepackage{epstopdf}
\usepackage{amsmath}
\usepackage{amsfonts}
\usepackage{color}
\usepackage{tabularx}
\usepackage{textcomp}

\begin{document}
\author{Jakub P\c ekalski}
\email{jpekalski@ichf.edu.pl}
\affiliation{Institute of Physical Chemistry, Polish Academy of Sciences, 01-224 Warszawa, Poland}
\author{Eldar Bildanov}
\affiliation{Belarusian State Technological University, Sverdlova 13a Minsk, Belarus}
\author{Alina Ciach}
\affiliation{Institute of Physical Chemistry, Polish Academy of Sciences, 01-224 Warszawa, Poland}
\title{Self-Assembly of Spiral Patterns in Confined System with Competing Interactions}

\begin{abstract}
Colloidal particles in polymer solutions and functionalized nanoparticles often exhibit short-range attraction coupled with long-range repulsion (SALR) leading to the spontaneous formation of symmetric patterns. Chiral nanostructures formed by \textcolor{black}{thin film} of SALR particles have not been reported yet.  In this study, we observe striking topological transitions from a symmetric pattern of concentric rings to a chiral structure of a spiral shape, \textcolor{black}{when the system is in hexagonal confinement}. We find that the spiral formation can be induced either by breaking the system symmetry with a wedge, or by melting of the rings. In the former case, the chirality of the spiral is determined by orientation of the wedge and thus can be controlled. In the latter, the spiral rises due to thermally induced defects and is absent in the average particle distribution that form highly regular hexagonal patterns in the central part of the system. These hexagonal patterns can be explained by {\it interference of planar density waves.} {\textcolor{black}{Thermodynamic considerations indicate that equilibrium spirals can appear spontaneously in any stripe-forming system confined in a hexagon with a small wedge, provided that certain conditions are satisfied by a set of phenomenological parameters.}}

\end{abstract}

\maketitle


Chiral two-dimensional (2d) nanostructures have appealing properties such as negative refractive 
index, optical rotatory dispersion or circular dichroism \cite{huang2004self,valev2009plasmonic}
that is useful e. g. in analysis of chirality in proteins or amino acids \cite{fasman2013circular}.
Chiral patterns can be obtained by different lithographic methods, lithography however has a 
drawback of the optical limit which makes the nanoscale world hardly accessible \cite{wan2015limits}.
Molecular self-assembly is a bottom-up approach, that is an alternative for lithography. 
\textcolor{black}{In 3d bulk it was shown both in experiment \cite{campbell:05:0} and simulations 
\cite{sciortino:05:0} that spherical hard-core particles which interact {\it via} competing short-range attractive and long-range
repulsive (SALR) forces can self-assemble into locally chiral structures. In the Letter we report on a quasi 2d SALR system which self-assembles into globally chiral structures with chirality that can be determined by the system geometry.}

\textcolor{black}{In 2d, computer simulations \cite{imperio:04:0,almarza:14:0}, theory \cite{pekalski:14:0,archer:08:0} and experiments \cite{ghezzi:97:0, sear:99:0, law:13:0} show that particles in SALR thin films can form patterns made of clusters, stripes or voids when their density increases.
In particular, on 2d interfaces SALR interactions were found between colloidal particles \cite{ghezzi:97:0,law:13:0} and between quantum dots \cite{sear:99:0}.
On the other hand, efforts to fabricate three-dimensional ordeblack structures with SALR particles have been so far in vain \cite{royall:18:0}}.  Importantly, 
the patterns pblackicted by the theory and simulations strongly resemble those 
obtained in diblock copolymer (BCP) thin films \cite{brassat:18:0, kim:15:0, ciach:13:0},
which found applications in variety of fields including filtration, photonics or
nanofabrication \cite{kuzyk:12:0, tzuang:09:0,pang:05:0}. Despite many similarities with BCPs, 
not much is known about properties of SALR patterns in confinement.
Our present study is aimed to fill this gap.

A single layer of a hexagonally ordeblack BCP forms stripped thin film that was broadly
studied and used for fabrication of e. g. nanoelectronics 
\cite{tsai2014two,kim:09:0,herr:11:0,schacher:12:0,liu2012epitaxial}.
SALR particles also form stripes when adsorbed at a two-dimensional interface, 
but experiments 
reported a number of topological defects that lead
to short-range ordering \cite{sear:99:0}. An order enhancement can be obtained by
imposing rigid boundary conditions provided that the geometry of the system fits
the symmetries of the pattern and that the distance between the walls is commensurate
with the period of the bulk pattern, $L_0$ \cite{choi:17:0}. In particular, a slit 
confinement allows for spontaneous formation of parallel and straight stripes, 
however, not fitting the wall-wall distance to $L_0$ results in stripe corrugations 
that spread along the whole system \cite{almarza:16:0}. Stripped SALR patterns on 
closed surfaces were only investigated in the case of particles adsorbed at the 
surface of a sphere \cite{zarragoicoechea:09:0, amazon:13:0,goh:13:0,amazon:14:0, pekalski:18:0},
where topological defects are enforced by the system topology. 
Here, we report on a closed system in which 
topological defects do not appear at low temperatures regardless of the system size,
and the boundary conditions lead to self-assembled chiral structures with desiblack chirality.

{\bf{Methods and model.}} \textcolor{black}{  To study confinement effects, one needs to know the
bulk behavior of the consideblack system. In the case of SALR particles, 
competing interactions lead to highly complex energy landscape, Kosterlitz-Thouless 
transitions, cluster formation, periodic ordering and thus sampling of its phase space
can be tricky \cite{imperio:04:0,archer:08:0}. Those difficulties are much easier 
to manage in a discretized space. For that reason the core of our methodology are 
Monte Carlo (MC) simulations of the lattice model studied in 
Refs.~\cite{almarza:14:0,pekalski:14:0,almarza:16:0}. Our 
main results, however, were verified by Molecular Dynamics simulations
for a continuous model. In what follows we briefly describe the lattice approach, 
but more detailed description of the lattice and the off-lattice simulations can be
found in the Supplemental Material (SM).}

  \begin{figure}[t]
      \includegraphics[scale=0.9]{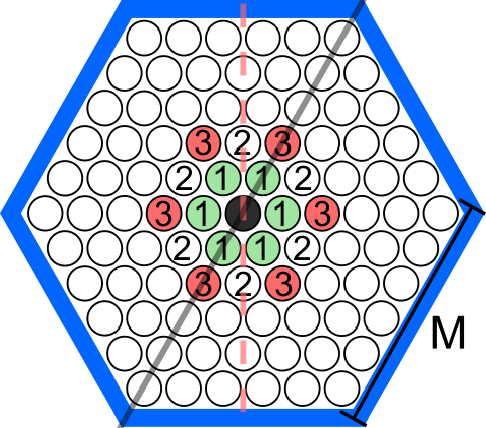}
      \caption{The hexagonal model with a single-occupancy triangular lattice. 
      The interparticle short-range attraction (SA) acts only between the first
      neighbors, while the long-range repulsion (LR) only between the third. 
      }
      \label{fig:model}
 \end{figure}

\textcolor{black}{We consider a triangular lattice model 
(Fig. \ref{fig:model})  with the interaction potential:}
\begin{equation}
\label{V}
V(\Delta{\bf  x}) = \left\{ \begin{array}{ll}
-J_1 & \textrm{for $|\Delta{\bf  x}| = 1$}, \;\; \;{\rm (nearest\; neighbors)}\\
+J_2 & \textrm{for $|\Delta{\bf  x}| = 2$}, \;\; \;{\rm (third \; neighbors)} \\
0 & \textrm{otherwise,}
\end{array} \right.
\end{equation}
where $-J_1$ and $J_2$ represent the attraction well and the repulsion barrier respectively.
The hard core repulsion between the particles was obtained by taking the particle diameter
$\sigma$ as a lattice constant. Similarly as in Refs. 
\cite{almarza:16:0,pekalski:15:0,pekalski:15:1,pekalski:14:0,almarza:14:0}
the repulsion to attraction ratio, $J_2/J_1$ was set to 3. With the above 
interaction potential the thermodynamic Hamiltonian has the form:
\begin{equation}
\label{H}
 H= \frac{1}{2}\sum_{\bf x} \sum_{\bf x'}\hat \rho({\bf x})V({\bf x}-{\bf x}')\hat \rho({\bf x'})
-\mu \sum_{\bf x} \hat \rho({\bf x}),
\end{equation}
where $\sum_{\bf x}$ denotes the summation over all lattice sites,
the microscopic density at the site ${\bf x}$ is $\hat\rho({\bf x})=1(0)$ when the site ${\bf x}$
is (is not) occupied and $\mu$ is the chemical potential. The results are presented in blackuced
units, i. e. $T^* = k_BT/J_1$, $\mu^* = \mu/J_1$.

In the bulk, periodic patterns
made of rhomboidal clusters, stripes or rhomboidal voids are formed depending on the value
of the chemical potential. In the current study we set $\mu^* =6$, corresponding to the stability of the stripped phase.

\begin{figure}[t]
    \begin{center}
	\centering
      \includegraphics[scale=0.10]{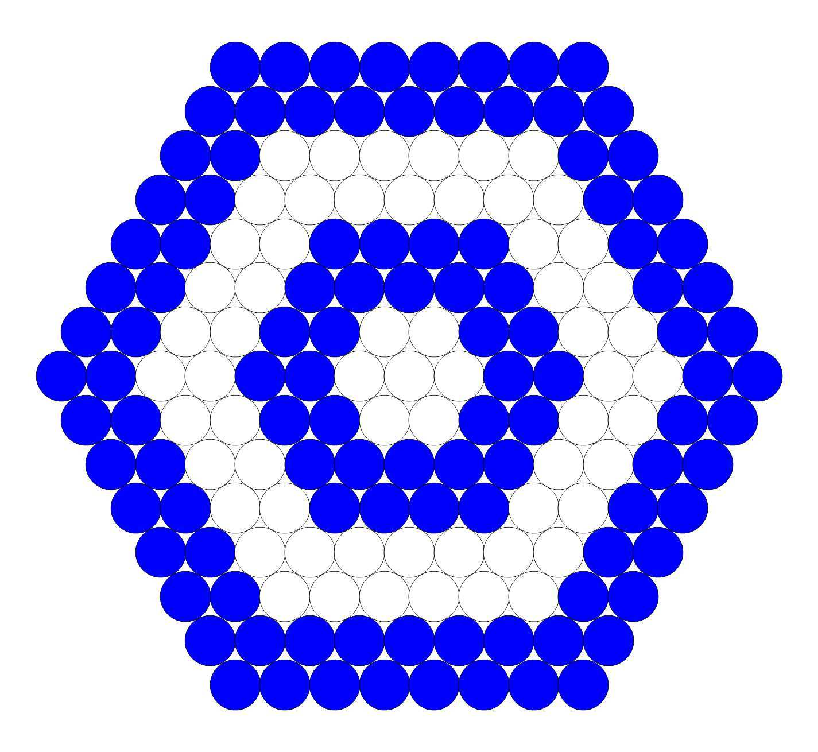}
      \includegraphics[scale=0.10]{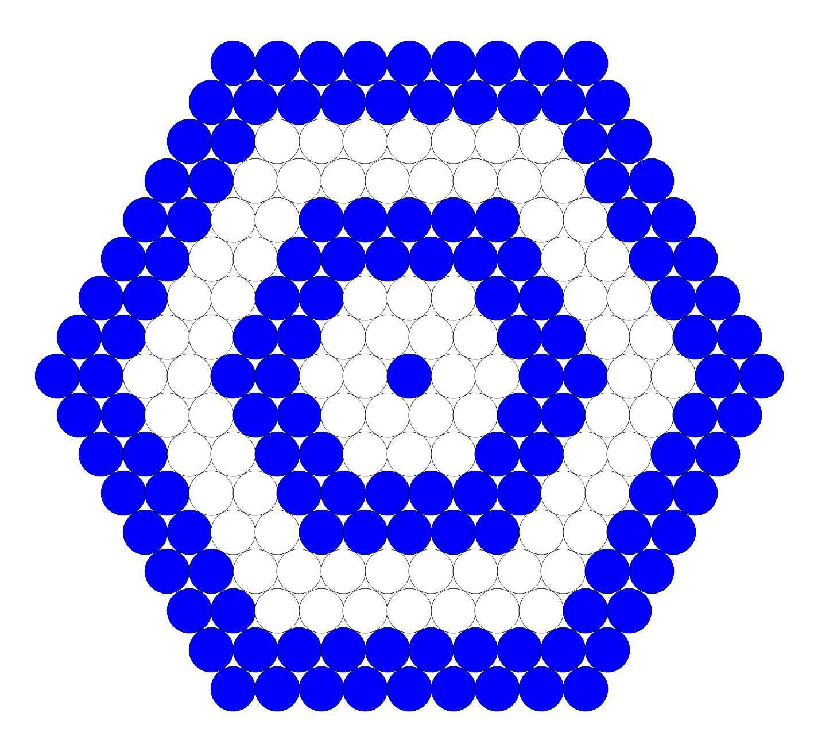}
      \includegraphics[scale=0.10]{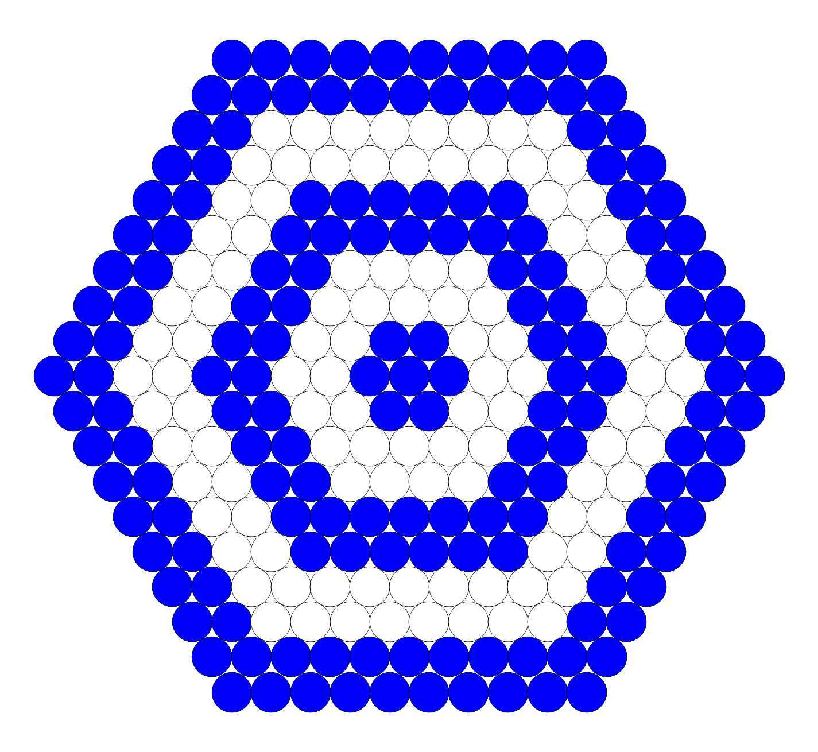}
      \includegraphics[scale=0.10]{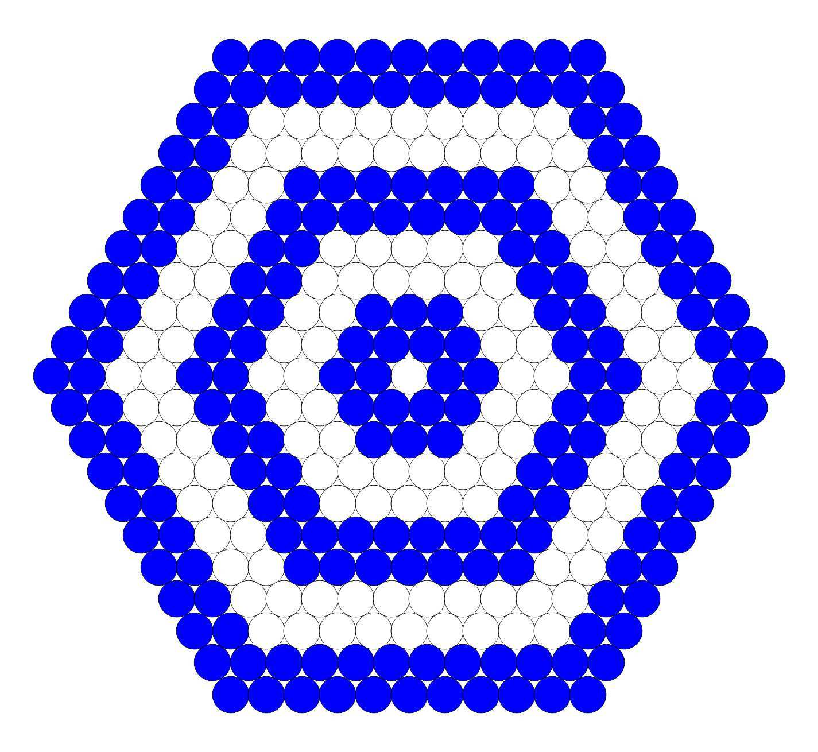}
      
      \caption{The ground state conformations for different system sizes, $M$. From left to right:
      $M = 8,9,10,11$.
      }
      \label{fig:gs_hex}
    \end{center}
\end{figure}

{\bf The ground state (T=0).} 
In the grand canonical ensemble, the stable structures are those which correspond to the 
global minimum of the grand potential per lattice site. At $T=0$, the grand potential 
blackuces to the thermodynamic Hamiltonian given by Eq. (\ref{H}). We found that for $\mu^*=6$ 
concentric rings are formed in the hexagonal confinement independently of
the system size
(Fig. \ref{fig:gs_hex}). 
The only difference
between the found structures is \textcolor{black}{the defect formed}
in the center of the hexagon: a hexagonal void, 
a single particle, a hexagonal cluster or a single empty site. Because the period of the bulk 
lamellar phase is 4, what determines the particle arrangement in the center of the hexagon
is the remainder, $r$, obtained when dividing $M$ by 4. Similar conformations were previously
described for BCP system within a circular template~\cite{choi:17:0}.

\begin{figure}[b]
    \begin{center}
	\centering
         \includegraphics[scale=0.4]{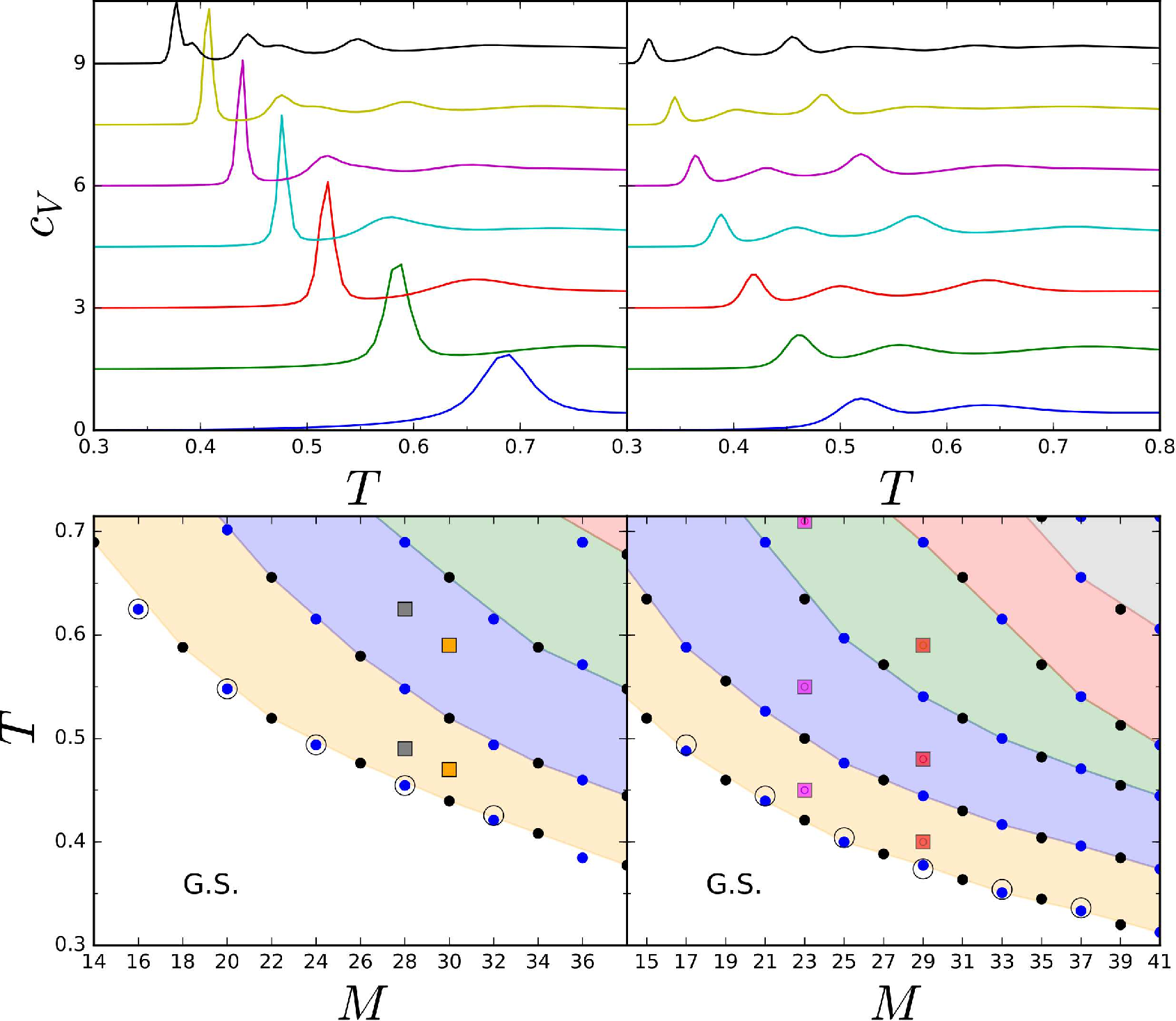}
      \caption{Upper panels: $c_V(T)$ for different sizes of the hexagon side 
      $M = 4 \times n + r$, vertically shifted for visibility. Left upper panel
      from the bottom to the top: $M=14,18,...36$, right upper panel from the 
      bottom to the top: $M=15,19,...,39$. Lower panels: temperatures corresponding 
      to the $c_v(T)$ peaks. G.S. indicate region for which the ground state structures 
      depicted on Fig. \ref{fig:gs_hex} are stable. Density distributions and snapshots 
      for $(M,T)$ values indicated by squares are shown on Fig. \ref{fig:maps}. The circles
      correspond to the temperature at which the average number of clusters, $c(T)$, drops significantly.
     \textcolor{black}{ The plot of $c(T)$ for different system sizes is shown in the SM.
      }}
      \label{fig:cv}
    \end{center}
\end{figure}

{\bf The temperature effect.} 
We start
our study of temperature effects on the hexagonal system by the analysis of the heat capacity, $c_V(T)$.
Fig. \ref{fig:cv} shows that the 
dependence of $c_V$ on $T$ is significantly different for $r = 0,2$ and $r=1,3$.
 \textcolor{black}{In the former case, low-T $c_V$ peaks are significantly higher and sharper,
and occur  at higher temperatures than for $r=1,3$ . 
 Also the number of 
peaks is different in the two cases. With increasing system size, new maxima rise 
and $c_V(T)$ becomes more complex. 
For each value of $r$, the temperature of the low-T peak of $c_V$, 
decreases monotonically,  but for all the consideblack system sizes is still larger than the bulk value, 
$T_{\text bulk}^* = 0.25$
that corresponds to the lamella - molten lamella phase transition~\cite{almarza:14:0}. }

\begin{figure}
	\centering
      \includegraphics[scale=1.0]{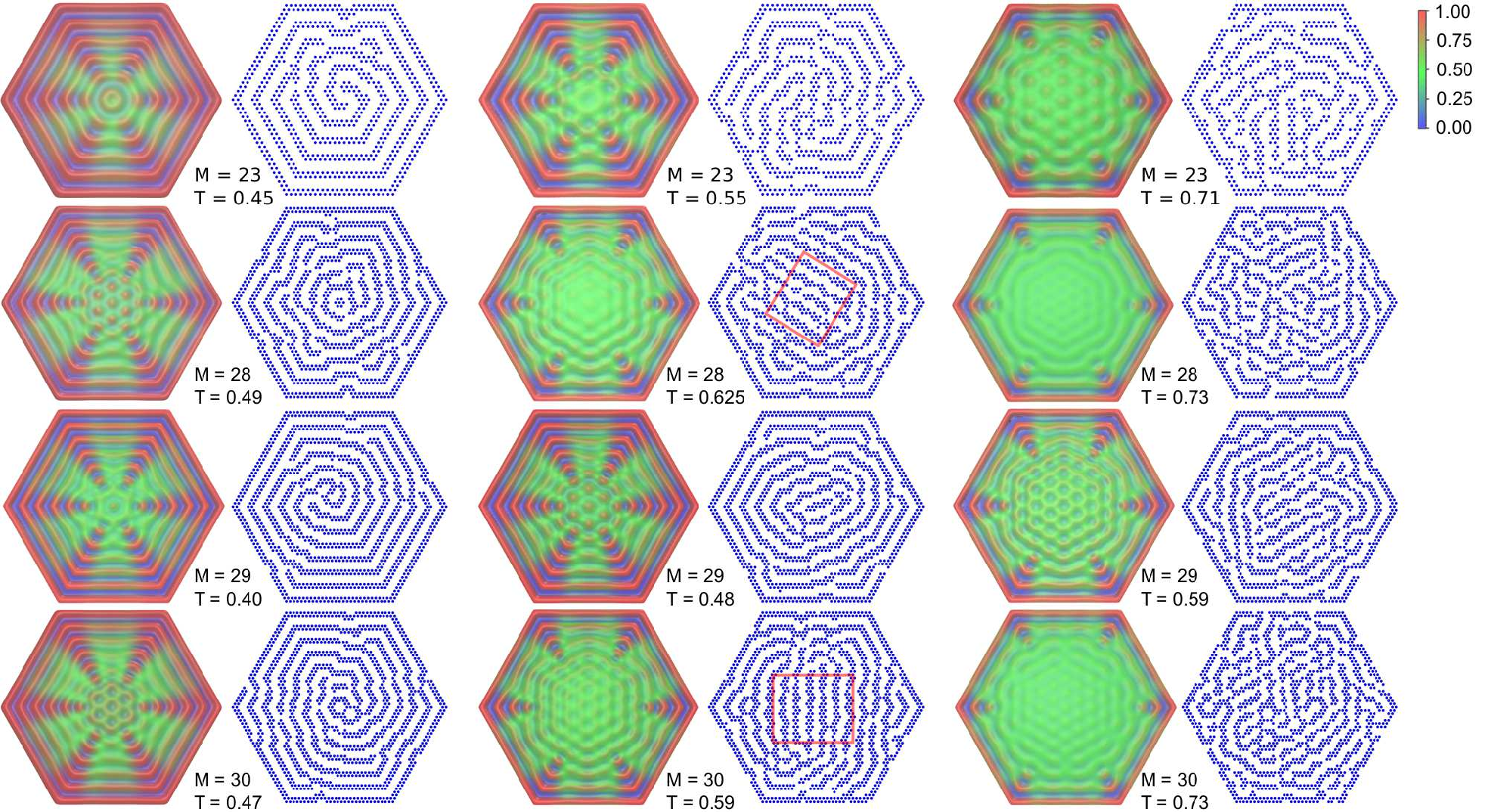}
      \caption{Average density distributions and the corresponding snapshots for different $(M,T)$ points marked by squares on Fig. \ref{fig:cv}. 
     In the snapshots, the particles are shown in blue. The thin black rectangle encloses a local lamellar structure that forms in the center.}
      \label{fig:maps}
\end{figure}

 The average density distributions and the representative snapshots are shown in
Fig. \ref{fig:maps} for temperatures corresponding to the minima of $c_V$.
 The average density distributions
are highly symmetric. Of course, the symmetry of the density distributions
follows from the symmetry of the hexagonal confinement. 
The snapshots are not symmetric and appear disordeblack. A closer look reveals that
near  the system boundaries the  stripes are parallel to them. In the central part of the system, 
however, stripes parallel to each other  and perpendicular to the diagonal of 
 the hexagon can be seen in several cases (see Fig.\ref{fig:maps}).
  Probabilities of configurations obtained by a rotation by the angles $\pi/3, 2\pi/3$ are the same. 
  Hence, averaging over all configurations  should have an effect similar to a superposition of standing
  planar density waves leading to a hexagonal pattern in the central part of the system. 
 \textcolor{black}{
 Such patterns should be present in many stripe forming systems in a 
 hexagonal confinement, and indeed were observed experimentally  \cite{ldos1,ldos2}.
  For different phase shifts of the interfering waves one can find centrally located structures that 
 resemble a honeycomb or a pattern of hexagonally ordeblack clusters. 
 Such patterns can be seen in Fig.\ref{fig:maps}
 for different values of $r$ and $T$}. The density profiles 
 computed along the symmetry axes of the hexagon show a minimum or a
 maximum in the center of the hexagon for $r=0,3$ or $r=1,2$ respectively (SM).

\begin{figure}[t]
    \begin{center}
	\centering
      \includegraphics[scale=0.9]{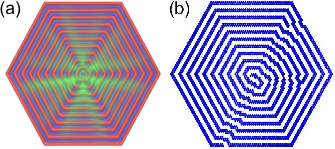}
      \caption{Density map (a) and a single configuration (b) of the SALR system with side of 
      size $M=41$ at $T=0.35$. The coloring scheme is the same as in Fig. \ref{fig:maps}.
      }
      \label{fig:spiralsym}
    \end{center}
    \begin{center}
	\centering
      \includegraphics[scale=0.9]{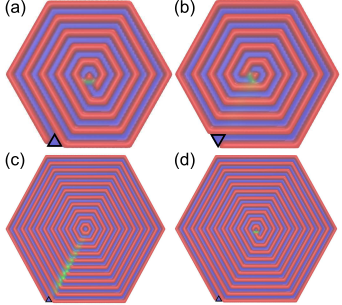}
      \caption{Low temperature density maps of the system with different system sizes. 
      Panels (a) and (b): $M=21$, panel (c): $M=44$, panel (d): $M=43$. The wedge position is marked by a black triangle and the color 
      code is the same as in Fig. \ref{fig:spiralsym}.
      }
      \label{fig:spiralwedge}
    \end{center}
\end{figure}

\textcolor{black}{In the bulk}, an increase of temperature results in breaking of the
lamellar segments into smaller pieces.  In the case of the hexagonal system, 
the rings break or
merge. For larger systems with $M>40$ and $r=1,3$, a one stripe of a spiral shape is formed,
\textcolor{black}{as can be seen in Fig.~\ref{fig:spiralsym}.}
Importantly, in the symmetric hexagonal system, the spiral can only be
seen on the level of a single configuration. The average density
distribution should have the same symmetries as the confinement and cannot be spiral. In what
follows, we show how to induce spiral formation and control its chirality.

{\bf Inducing a spiral.} 
In order to check how a spiral configuration can be induced 
by the confinement, we tried to break the system symmetry in several ways. We found that
introducing a triangular wedge can trigger formation of a stable spiral,
as long as the side of the wedge is  equal to the period of the striped structure
(Fig. \ref{fig:spiralwedge}ab). What is more, the chirality 
of the spiral can be easily controlled by positioning of the wedge.
It turns out, however that formation of a defectless spiral 
depends on the system size, and happens only for $r = 1,3$ (Fig. \ref{fig:spiralwedge}bc).

{\bf  \textcolor{black}{Thermodynamic considerations}}

 \textcolor{black}{
In what follows we develop a simple thermodynamic description of a quasi-two dimensional stripe-forming
system in hexagonal confinement. Our goal is to find out in what thermodynamic conditions
the self-assembly into concentric rings if 
the system is hexagonal, and into a spiral pattern if the system has a symmetry broken by 
a triangular wedge can occur, when the stripped pattern is formed in the unrestricted system.}

 \textcolor{black}{
The stable structures are determined by the grand thermodynamic potential $\Omega$, 
which is a sum of the contributions proportional to the area of the hexagon, $A$,
the side length, $M$, and a contribution associated with point-like defects, $D$. Thus, we postulate 
\begin{equation}
 \Omega=\omega_bA +(\Gamma_w+\Gamma_{db})M+D
\end{equation}
where $\omega_b$ is the grand potential per surface area in the unconfined system,
$\Gamma_w$ is the sum of the surface tensions at all the confining walls,
and $\Gamma_{db}$ is the sum of the surface tensions at the boundaries between domains with different stripe-orientation.
}
 \textcolor{black}{
Inside a symmetric hexagon one can expect ordeblack states with either parallel stripes or concentric 
rings. In the former case,  $\Gamma_{db}=D=0$ and
$\Gamma_w=2\gamma_{\parallel}+4\gamma_t$, where $\gamma_{\parallel}, 
\gamma_t$ denote the surface tension for stripes parallel 
or tilted to the walls, respectively. For concentric rings $\Gamma_w=6\gamma_{\parallel}$,  $\Gamma_{db}=6\gamma_{db}$,
and $D=d_c$, where $d_c$ comes from the defect in the center, \textcolor{black}{ and depends on $M$ (Fig.\ref{fig:gs_hex})}.
Thus, the stable structure is determined by phenomenological parameters such as $\gamma_{db}$ 
which is associated with bending rigidity of the stripes. The concentric rings are stable if $4(\gamma_t-\gamma_{\parallel})-6\gamma_{db} -d_c/M>0$, i.e. when the stripes parallel to the walls are much more favorable than
the tilted ones, and the free-energy cost of bending of the stripes is small. 
}

 \textcolor{black}{
When a triangular wedge is introduced at a vertex of the hexagon, then the ring adsorbed at
the walls breaks, and its two ends associated with the excess energy $d_e$ occur at 
the two sides of the wedge. Breaking of the stripe leads to $D=d_c+2d_e$. 
If a spiral instead of concentric rings appears,  \textcolor{black}{
then the length of the domain-boundary initiated by the wedge shortens leading to
a negative contribution to $D$. Furthermore, a stripe end appears
in the center of the hexagon, leading to another contribution to $D$.
 The  sum of the two contributions to $D$ is denoted by $d_s$.}
 Thus, the grand potential difference between the spiral and the concentric hexagons is 
$ (d_e + d_s) - (2d_e +d_c) = d_s-d_e-d_c$. If $d_s-d_e-d_c<0$,
the spiral is more stable.
}

 \textcolor{black}{
 Our thermodynamic considerations indicate that there may exist many systems forming 
 chiral patterns on different length scales,
provided that the phenomenological parameters satisify the inequalities 
$4(\gamma_t-\gamma_{\parallel})-6\gamma_{db} -d_c/M>0$ and $d_s-d_e-d_c<0$.
In our model spirals are formed
for $r=1,3$, and are not formed for $r=0,2$, indicating a smaller value of $d_c$ in the latter case (see SM for the calculation details).
}

{\bf Conclusions.} We have analyzed the self-assembly of \textcolor{black}{a  stripe forming system} when exposed to spatial confinement of hexagonal shape. Our results reveal two possible scenarios for obtaining a spiral structure: (i) by a proper temperature adjustment or (ii)
by introducing a wedge that breaks the symmetry of the system and determines the location of the spirals terminal point. In the former case, the temperature controls the number of topological defects and thus the self-assembled stripes can \textcolor{black}{break and} merge into one stripe of a spiral shape. However, neither the spiral orientation nor the position of the terminal point can be controlled if the system is symmetric. Introducing the wedge gives that control over the spiral. Importantly, the size of the triangular wedge requiblack for inducing the spiral, does not depend on the system size, but is determined by the stripe width. Thus, a relatively small obstacle on the confining wall can lead to striking topological changes in particles conformation.

Our results were obtained for the SALR system in which isotropic competing interactions
lead to formation of stripes. The proposed model-free thermodynamic approach shows, 
however \textcolor{black}{ 
that for many stripe forming  systems the
hexagonal confinement may favor formation of concentric rings if the system is symmetric, 
or spirals if the symmetry of the system is broken by a wedge of specific size. 
These structures can be stable when thermodynamic parameters, such as the surface tensions and the free-energies of defects, obey certain inequalities.}
Importantly, such general conclusion finds support in previously conducted experimental studies
on BCPs \cite{choi:17:0}.
 
The analysis of both thermal and structural properties of the system indicates that the hexagonal 
confinement enhances the integrity of stripes. The melting temperature at which concentric rings
start to merge is significantly higher than the stripes melting temperature in the bulk. On 
the other hand, above the melting point defects induced by the temperature can destroy the ordering,
so that the average density distributions and single snapshots are tremendously different.
Thus, the obtained density maps, although highly symmetric, do not indicate formation of
ordeblack patterns \textcolor{black}{unless the time of observation is much larger than
the time scale characterizing the dynamics of the system}.
\textcolor{black}{Interestingly, similar patterns were observed on the quantum level, 
where scattering of surface state electrons at the edges of densely packed Au atoms, 
lead to formation of standing waves by the local density of states \cite{ldos1,ldos2}. 
Thus, the unexpected family of highly symmetric patterns shown in Fig. \ref{fig:maps} 
can actually be present at length scales ranging from few nm to hundblacks of micrometers.}


\section{Acknowledgements}
The authors would like to thank Wojciech T. G\'o\'zd\'z for several enlightening discussions during the course of this work.

This project has received funding from the European Union\textquotesingle s Horizon 2020 research and innovation programme under the Marie
Sk\l{}odowska-Curie grant agreement No 734276 (CONIN). An additional support in the years 2017-2018  has been granted  for the CONIN project by the Polish Ministry of Science and Higher Education. Financial support from the National Science Center under grant No. 2015/19/B/ST3/03122 is also acknowledged.


\end{document}